\RequirePackage{fix-cm}
\documentclass[10pt,a4paper,twocolumn]{scrartcl}
\usepackage[utf8]{inputenc}
\usepackage[T1]{fontenc}
\usepackage{geometry}
\usepackage{textcomp}
\usepackage{prettyref}
\usepackage{graphicx}
\usepackage[hidelinks]{hyperref}
\usepackage{authblk}
\usepackage{import}
\usepackage{tikz}
\usepackage{tikz-timing}
\usepackage{xifthen}
\usepackage{listings}
\usepackage[binary-units]{siunitx}
\usepackage{amsmath}
\usepackage{listings}
\usepackage[xindy]{glossaries}
\usepackage{hyphenat}

\usepackage[EULERGREEK]{sansmath}

\usepackage[colorinlistoftodos,prependcaption,textsize=tiny]{todonotes}
\usetikzlibrary{calc,shapes,fit,positioning,arrows}

\newrefformat{fig}{Fig.~\ref{#1}}
\newrefformat{tab}{Table~\ref{#1}}

\definecolor{red}{HTML}{AF5A50}
\definecolor{blue}{HTML}{005B82}
\definecolor{green}{HTML}{7D966E}
\definecolor{yellow}{HTML}{D7AA50}

\lstdefinestyle{tclstyle}{
  language   = tcl,
  basicstyle = \ttfamily\scriptsize,
	basewidth  = {.5em,0.4em},
	columns    = flexible,
  keepspaces = true,
}
\lstdefinestyle{pythonstyle}{
  language   = python,
  basicstyle = \ttfamily,
	basewidth  = {.5em,0.4em},
	columns    = flexible,
  keepspaces = true,
}

\setcapindent{0pt}

\deffootnote[0.3em]{0.0em}{0em}{\textsuperscript{\thefootnotemark}}

\makeatletter
\patchcmd{\@maketitle}{\LARGE \@title}{\fontsize{16}{19.2}\selectfont\@title}{}{}
\makeatother

\newacronym{ppu}{PPU}{plasticity processing unit}
\newacronym{padi}{PADI}{parallel driver interface}

\newacronym{sram}{SRAM}{static random-access memory}
\newacronym{adc}{ADC}{analog-to-digital converter}
\newacronym{dac}{DAC}{digital-to-analog converter}
\newacronym{dut}{DUT}{design under testing}
\newacronym{asic}{ASIC}{application-specific integrated circuit}
\newacronym{fpga}{FPGA}{field-programmable gate array}
\newacronym{vlsi}{VLSI}{very-large-scale integration}
\newacronym{simd}{SIMD}{single instruction, multiple data}
\newacronym{ocp}{OCP}{Open Core Protocol}
\newacronym{soc}{SoC}{system on a chip}
\newacronym{sta}{STA}{static timing analysis}
\newacronym{gals}{GALS}{globally asynchronous locally synchronous}
\newacronym{dpi}{DPI}{SystemVerilog direct programming interface}
\newacronym{pll}{PLL}{phase-locked loop}
\newacronym{fifo}{FIFO}{First-In-First-Out Buffer}

\newacronym{psc}{PSC}{post-synaptic current}
\newacronym{psp}{PSP}{post-synaptic potential}
\newacronym{stp}{STP}{short-term plasticity}
\newacronym{stdp}{STDP}{spike-timing-dependent plasticity}
\newacronym{rstdp}{R-STDP}{reward-modulated spike-timing-dependent plasticity}
\newacronym{lif}{LIF}{leaky integrate-and-fire}
\newacronym{adex}{AdEx}{adaptive exponential leaky integrate-and-fire}

\newacronym{mc}{MC}{Monte Carlo}

\title{Verification and Design Methods for the BrainScaleS Neuromorphic Hardware System}

\date{11th October 2019}

\begin{document}

\author[$\star$,1]{Andreas Grübl}
\author[$\star$,1]{Sebastian Billaudelle}

\author[1]{\\Benjamin Cramer}
\author[1]{Vitali Karasenko}
\author[1]{Johannes Schemmel}

\affil[$\star$]{Authors with equal contribution}
\affil[1]{Kirchhoff-Institute for Physics, Heidelberg University}

\maketitle

\begin{abstract}
\noindent\bfseries
This paper presents verification and implementation methods that have been developed for the design of the BrainScaleS-2 \SI{65}{nm} ASICs.
The 2nd generation BrainScaleS chips are mixed-signal devices with tight coupling between full-custom analog neuromorphic circuits and two general purpose microprocessors (PPU) with SIMD extension for on-chip learning and plasticity.
Simulation methods for automated analysis and pre\hyp{}tapeout calibration of the highly parameterizable analog neuron and synapse circuits and for hardware\hyp{}software co\hyp{}development of the digital logic and software stack are presented.
Accelerated operation of neuromorphic circuits and highly-parallel digital data buses between the full-custom neuromorphic part and the PPU require custom methodologies to close the digital signal timing at the interfaces.
Novel extensions to the standard digital physical implementation design flow are highlighted.
We present early results from the first full-size BrainScaleS-2 ASIC containing 512 neurons and 130\,K synapses, demonstrating the successful application of these methods.
An application example illustrates the full functionality of the BrainScaleS-2 hybrid plasticity architecture.

\end{abstract}
\vspace{0.5em}
{
\bfseries
\itshape
\noindent
neuromorphic hardware, plasticity, mixed-signal, verification, physical design, methodology
}

\glsresetall
\section{Introduction}
\label{sec:intro}
The design of neuromorphic hardware follows the goal to model parts, or at least functional aspects, of the biological nervous system.
A main motivation is to reproduce its computational functionality and especially its ability to efficiently solve cognitive and perceptual tasks. %
Achieving this requires modeling networks of a sufficient complexity in terms of number of neurons and number of synaptic connections.
The brain as a whole and especially its ability to learn and adapt to specific problems is still subject to basic neuroscientific research. %
Consequently, flexible implementations of learning and plasticity are desirable as well.

Several neuromorphic hardware systems have been proposed and developed that differentiate themselves in terms of architecture, scaling and learning capabilities, and whether they follow an analog/mixed-signal or purely digital approach.
TrueNorth \cite{merolla2014million} is a neuromorphic chip that integrates 4096 neurosynaptic cores to simulate 1\,M neurons and 256\,M synaptic connections at biological real-time. %
It is fully digital and the cores are operated asynchronously. Learning algorithms need to be implemented off-chip; multi-chip topologies have been proposed in \cite{akopyan2015truenorthFlow}.
The SpiNNaker system \cite{furber2014spinnaker} is based on processor nodes comprising 18 ARM cores which are interconnected using an asynchronous networking infrastructure, optimized for the high-fanout routing of neural events. %
It provides the flexibility to change the underlying simulation models in software and is designed to operate at biological real-time, while simulation speed might decrease with increased model complexity. %
It provides 1\,M cores in its current state.
Online learning can be implemented in software, which also results in a performance penalty \cite{diehl2014spinnaker_stdp}. %
Intel's Loihi chip \cite{davies2018loihi} contains 128 neuromorphic cores and is capable of simulating 130\,k neurons and 130\,M synapses in real-time.
It features on-chip learning capabilities that allow for different types of synaptic plasticity.
There exists a multichip platform containing 64 Loihi chips. %
While the aforementioned systems are implemented using digital logic, analog neuromorphic designs make use of dedicated analog circuits as computational elements, which is beneficial in terms of energy and cost efficiency %
and their continuous-time operation reproduces the collective dynamics of neural networks more faithfully.
One recent example is the Dynap-SEL chip \cite{moradi2018dynapsel} which comprises 1.1\,k neurons and 78\,k synapses, which are partially capable of on-chip learning using \gls{stdp}.
An in-depth review of a selection of current analog neuromorphic hardware systems can be found in \cite{spikingArrays2018}, a general overview in \cite{Furber_2016}.

In this paper we describe aspects of the familiy of BrainScaleS systems %
which similarly aim at providing a tool for neuroscientists to facilitate large-scale spiking neural network emulations at a sufficiently high level of biological detail.
Instead of integrating model equations numerically we implement a physical system using analog circuits which can be described by the same equations. 
The model variables evolve in continuous-time, determined by the circuit parameters.
Quantities like reversal potentials, currents, or conductances can directly be translated to our %
circuits. Membrane and synaptic time constants also follow from the mapping of the equation dynamics. 
In the BrainScaleS systems we selected the circuit elements in a way that these characteristic times are shorter than in biology.
As a consequence, the physical model operates at a speedup of $10^3$ to $10^4$ compared to biological time scales. %
BrainScaleS-1 introduced wafer-scale integration to allow for the emulation of networks of up to 200\,K neurons and 44\,M synapses on a single silicon wafer \cite{schemmel2010iscas}.

The second version BrainScaleS systems transition from a $\SI{180}{\nano\meter}$ CMOS process to a $\SI{65}{\nano\meter}$ process node.
While the gain in available silicon area has mainly been used to add features to the analog circuits and improve their debugging capabilities and robustness \cite{aamir2018lifarray}, %
we could substantially increase the complexity of the surrounding digital logic.
The most prominent addition is a hybrid plasticity scheme, where learning algorithms can be freely programmed in software and executed on an embedded microprocessor, in contrast to the \gls{stdp}-based fixed learning algorithms in BrainScaleS-1 \cite{schemmel_ijcnn06}.
The processor is directly attached to the analog neuromorphic circuits \cite{friedmann2016hybridlearning}. %
Together with the sped-up operation of the analog circuits, this tight coupling requires high throughput and thus high operating clock frequency
and wide data paths in the digital logic.
This results in complex mixed-signal interfaces with multiple closed loops between analog and digital domains.

Complex interfaces, highly integrated analog circuit arrays combined with the difficulties and pitfalls of physical standard-cell design can push standard tooling to its limits.
This might be a common denominator of most neuromorphic hardware designs.
The development of non-standard design flows or custom tools is therefore an integral part of the overall design process. %
For TrueNorth and Loihi these strategies have been outlined
in \cite{akopyan2015truenorthFlow} and \cite{davies2018loihi}, respectively.
Such information is otherwise only sparsely available. %

To improve on this situation, this paper describes selected aspects of the implementation and verification strategies employed in the design of different version of the BrainScaleS-2 neuromorphic chips.
Our verification approach is described in \prettyref{sec:ms-if}.
First measurement results from the full-size BrainScaleS-2 chip containing 512 neuron and 130\,K synapse circuits demonstrate their successful application.
Non-standard implementation methodologies, especially for the tight coupling of large and dense analog arrays to comparably high-speed digital logic, are explained in \prettyref{sec:phys-impl}.
To illustrate the viability of the presented methodologies, \prettyref{sec:applications} presents results from the manufactured silicon by means of a reinforcement learning experiment that is executed on a BrainScaleS-2 chip.

\section{BrainScaleS Architecture}
\label{sec:architecture}

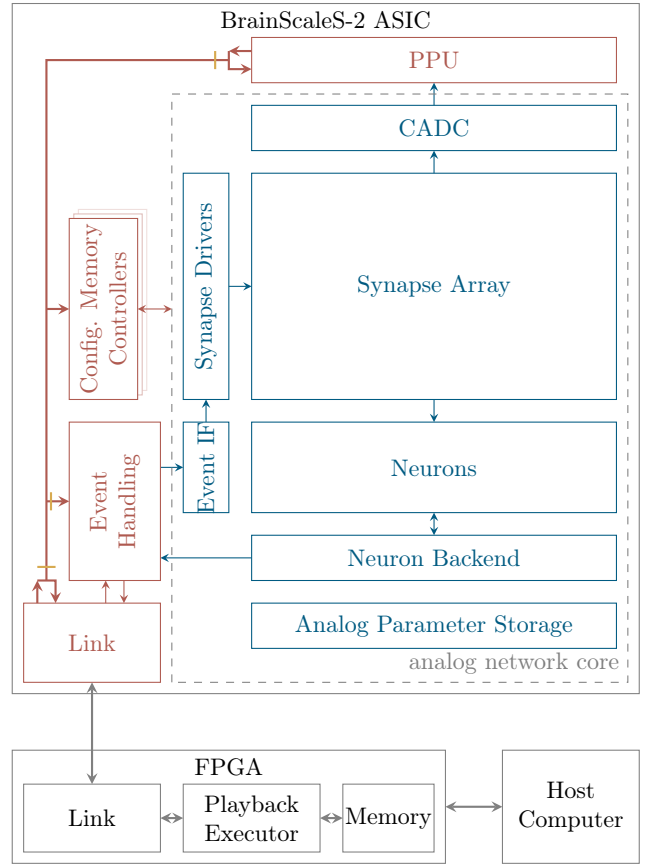
\begin{figure}
	\begin{center}
		\begin{tikzpicture}[scale=.6,every node/.style={scale=0.9}]
	\draw[blue] (1.5,3.0) rectangle ++(1,2) node[pos=0.5,rotate=90] {Event IF};
	\draw[blue,stealth-] ( 1.5,4.0) -- ++(-0.5,0.0);
	\draw[blue,-stealth] ( 2.0,5.0) -- ++(0.0,0.5);

	\draw[blue] (1.5, 5.5) rectangle ++(1,5) node[pos=0.5,rotate=90] {Synapse Drivers};
	\draw[blue,-stealth] (2.5,8.0) -- ++(0.5,0.0);
	
	\draw[blue] (  3, 0.0) rectangle ++(8,1) node[pos=0.5] {Analog Parameter Storage};
	
	\draw[blue] ( 3, 1.5) rectangle ++(8,1) node[pos=0.5] {Neuron Backend};
	\draw[blue,-stealth] ( 3.0,2.0) -- ++(-2.0,0.0);

	\draw[blue] ( 3, 3.0) rectangle ++(8,2) node[pos=0.5] {Neurons};
	\draw[blue,stealth-stealth] (7.0,3.0) -- ++(0.0,-0.5);

	\draw[blue] ( 3, 5.5) rectangle ++(8,5) node[pos=0.5] {Synapse Array};
	\draw[blue,-stealth] (7.0,5.5) -- ++(0.0,-0.5);
	\draw[blue] ( 3,11.0) rectangle ++(8,1) node[pos=0.5] {CADC};
	\draw[blue,-stealth] (7.0,10.5) -- ++(0.0,0.5);

	\draw[red] (3.0,12.5) rectangle ++(8,1) node[pos=0.5] {PPU};
	\draw[blue,-stealth] (7.0,12.0) -- ++(0.0,0.5);

	\draw[gray,dashed] (1.25,-0.75) rectangle ++(10.0,13.0);
	\node[gray,anchor=south east] at (11.25,-0.75) {analog network core};

	\draw[red] (-2.0,-0.75) rectangle ++(3.0,1.75) node[pos=0.5,text width=2cm,align=center] {Link};
	\draw[red,-stealth] (-0.0,1.0) ++ (-0.2,0.0) -- ++(0.0,0.5);
	\draw[red,stealth-] (-0.0,1.0) ++ (+0.2,0.0) -- ++(0.0,0.5);
	
	\draw[red] (-1.0, 1.5) rectangle ++(2.0,3.5) node[pos=0.5,text width=2cm,align=center,rotate=90] {Event\\Handling};

	\draw[red!020,fill=white] (-1.0,5.5) ++ (0.2,0.2) rectangle ++(1.5,4.0) node[pos=0.5,rotate=90,align=center] {};
	\draw[red!040,fill=white] (-1.0,5.5) ++ (0.1,0.1) rectangle ++(1.5,4.0) node[pos=0.5,rotate=90,align=center] {};
	\draw[red!100,fill=white] (-1.0,5.5) ++ (0.0,0.0) rectangle ++(1.5,4.0) node[pos=0.5,rotate=90,align=center] {Config. Memory\\ Controllers};
	\draw[red,stealth-stealth] (0.5,7.5) -- ++(0.75,0.0);

	\draw[red,thick] (-1.5,1.5) -- (-1.5,13.0) -- (2.5,13.0);
	
	\draw[red,thick,-stealth] (2.5,13.2) -- ++(-0.0,-0.4) -- ++(0.5,0.0);
	\draw[red,thick,stealth-] (2.5,13.2) -- ++(0.5,0.0);
	\draw[yellow,thick] (2.5,13.0) ++ (-0.3,-0.2) -- ++(0.0,0.4);
	
	\draw[red,thick,-stealth] (-1.5,1.5) ++ (-0.2,0.0) -- ++(0.4,0.0) -- ++(0.0,-0.5);
	\draw[red,thick,stealth-] (-1.5,1.5) ++ (-0.2,0.0) -- ++(0.0,-0.5);
	\draw[yellow,thick] (-1.5,1.5) ++ (-0.2,0.3) -- ++(0.4,0.0);

	\draw[red,thick,-stealth] (-1.5,7.50) -- ++(0.5,0.0);
	\draw[red,thick,-stealth] (-1.5,3.25) -- ++(0.5,0.0);
	\draw[yellow,thick] (-1.0,3.25) ++ (-0.4,-0.2) -- ++(0.0,0.4);
	
	\draw[draw=gray] (-2.25,-1.0) rectangle ++(13.75,15.25);
	\node[below] at (4.625,14.25) {BrainScaleS-2 ASIC};

	\draw[draw=gray] (-2.00,-4.50) rectangle ++(3.0,1.5) node[pos=0.5,text width=2cm,align=center] {Link};
	\draw[draw=gray] ( 1.50,-4.50) rectangle ++(3.0,1.5) node[pos=0.5,text width=2cm,align=center] {Playback Executor};
	\draw[draw=gray] ( 5.00,-4.50) rectangle ++(2.0,1.5) node[pos=0.5,text width=2cm,align=center] {Memory};
	\draw[draw=gray] (-2.25,-4.75) rectangle ++(9.5, 2.5);
	\node[below] at (2.5,-2.25) {FPGA};

	\draw[gray,thick,stealth-stealth] (-0.5,-3.00) -- ++(0.0,2.25);
	\draw[gray,thick,stealth-stealth] ( 1.0,-3.75) -- ++(0.5,0.00);
	\draw[gray,thick,stealth-stealth] ( 4.5,-3.75) -- ++(0.5,0.00);
	
	\draw[gray,thick,stealth-stealth] ( 7.25,-3.5) -- ++(1.25,0.00);

	\draw[draw=gray] ( 8.5,-4.75) rectangle ++(3.0,2.5) node[pos=0.5,text width=2cm,align=center] {Host\\ Computer};
\end{tikzpicture}
	\end{center}
	\caption{Block-level diagram of a BrainScaleS-2 system, including the \gls{asic} itself as well as an \gls{fpga} managing the communication to the host system.}
	\label{fig:architecture}
\end{figure}

The structure of the BrainScaleS-2 system is depicted in \prettyref{fig:architecture}.
The mixed-signal BrainScaleS-2 \gls{asic} contains \gls{vlsi} analog neuromorphic circuits, digital control and communication infrastructure, and one or more general-purpose microprocessors mainly intended to be used as \glspl{ppu}.
The \gls{asic} is implemented using a digital top-level description in a way that all analog signals are confined within the \gls{asic} and off-chip communication is carried out utilizing digital high-speed serial communication techniques. %
Real time experiment control is performed by an \gls{fpga} which also manages host communication.
Technical details relevant for this publication will be described in the following subsections.

\subsection{Analog neural network core}
\label{sec:anncore}

\paragraph{Synapse drivers}
\label{sec:anncore_syndrv}
The digital event handling logic injects events into a custom CMOS-level bus to distribute the spike events across an array of synapse drivers.
This event interface bus allows to target single or multiple synaptic rows by means of a row select address, which can be partially masked by the receiver circuits.
The signals on the event interface, depicted in \prettyref{fig:padi_timing}, allow the synapse drivers to derive timing signals for the synaptic \glspl{dac}, which are then driven across the synaptic rows by a bank of buffers \cite{friedmann2016hybridlearning}.

Synapse drivers also implement short-term synaptic plasticity \cite{tsodyks97neural}, also following a pre-synaptic implementation approach as in previous generations \cite{schemmel_iscas07}.
Virtual neurotransmitter levels are represented as voltages on storage capacitors.
Based on these voltage levels the length of the synaptic current pulse transmitted to the neuron is modified, resulting in a change of synaptic efficacy.

\paragraph{Synapse array}
\label{sec:anncore_synapse}
The main purpose of a synapse is to generate a synaptic current according to the timing provided by the synapse drivers -- and naturally their pre-programmed weight value.
This \SI{6}{\bit} weight is stored alongside a \SI{6}{\bit} address that is matched against an incoming event's address in local \gls{sram}.

In order to allow for STDP-derived learning rules, the synapse circuits also implement a local, analog circuit for measuring the correlation of pre- and post-synaptic spikes.
These correlation traces are stored on capacitors to be digitized for hybrid plasticity \cite{friedmann2016hybridlearning}, which is described in \prettyref{sec:digital_plasticity}.

\paragraph{Neuron circuits}
\label{sec:anncore_neuron}
Synaptic currents are forwarded to the neuron circuits \cite{aamir17dls3neuron}.
In both BrainScaleS generations they implement the adaptive exponential leaky integrate-and-fire model \cite{gerstner2009adex}
\begin{align*}
	\small
	C\frac{\text{d}V}{\text{d}t} &=-g_\text{L}(V-E_\text{L})+g_\text{L}\Delta_\text{T}\exp \frac{V-V_\text{T}}{\Delta_\text{T}} -w+I \,, \\
	\tau_w\frac{\text{d}w}{\text{d}t} &= a(V-E_\text{L})-w \,,
\end{align*}
which adds an adaptation state variable $w$ as well as an exponential non-linearity to the underlying LIF equation for the membrane potential $V$.
The neuron's operating point is determined by its membrane capacitance $C$, leak conductance $g_\text{L}$, reversal potential $E_\text{L}$, a soft threshold $V_\text{T}$, and an exponential slope $\Delta_\text{T}$; the strength of the adaptation current is determined by conductance $a$ and a spike-triggered increment $w \leftarrow w + b$.
Besides these differential equations, the model includes a spiking condition where a neuron emits an event as soon as its membrane voltage crosses a threshold.
In the neuromorphic implementation, these spikes are latched by the neuron's full-custom digital backend circuit, where a priority-encoder is used to arbitrate between and then digitize events from groups of neurons.
The events are then streamed out to the digital control logic.
Based on received events, the backend circuits also generate refractory timing and other auxiliary signals for the analog neuron implementation.

Each neuron instance is individually parameterizable using a massively integrated analog parameter storage \cite{hock13analogmemory}.
Besides 8 neuronal voltages and 16 currents, this \emph{capacitive memory} also also provides global parameters to other analog circuits within the analog core.

\subsection{Hybrid plasticity}
\label{sec:digital_plasticity}

BrainScaleS-1, like its predecessor Spikey, already featured an implementation of \gls{stdp} \cite{schemmel_ijcnn06,schemmel2010iscas}.
To allow for the execution of a wider range of plasticity algorithms, BrainScaleS-2 introduced a customly developed and freely programmable processing element (\gls{ppu} \cite{friedmann2016hybridlearning,githubnux}).
The custom general purpose core implements the Power ISA \cite{powerisa_203}.
It is accompanied by a \gls{simd} vector unit which is tightly coupled to the columnar interfaces of the analog network core.
This most notably includes a full-custom \gls{sram} controller to access the synaptic memory.
For the integration of correlation traces or membrane voltages into plasticity algorithms, a column-parallel single-slope analog-to-digital converter (CADC) is used to digitize these analog observables.
Additionally, the \gls{ppu} can access all other on-chip components via an on-chip bus fabric, described in \prettyref{sec:digital}.
This allows to incorporate e.g. neuronal firing rates as observables for plasticity algorithms and reconfigure on-chip components.

\subsection{Digital control}
\label{sec:digital}
On-chip communication is facilitated by a custom\hyp{}developed bus architecture, which implements a subset of the \gls{ocp} \cite{ocp30,githubomnibus}; it is illustrated with red arrows in \prettyref{fig:architecture}.
Both, the host the chip is attached to and the \glspl{ppu} can access the bus via its multi-master capabilities.
All configuration and control registers are connected to the bus.
It also interfaces with a number of \gls{sram} controllers for the analog core's full-custom configuration memory.
The design is organized in a \gls{gals} fashion:
the \glspl{ppu} each run in a separate clock domain with globally tunable clock frequency to trade off optimal performance and energy efficiency for a given task.
Likewise runs the on-chip bus in a dedicated clock domain together with all memory control and auxiliary logic.
As an exception to the \gls{gals} architecture, link and event handling are kept in a single clock domain to avoid jitter in event transport when passing through asynchronous FIFOs.
All clock domains are decoupled using asynchronous FIFOs, denoted by strokes across the arrows representing the on-chip bus in \prettyref{fig:architecture}.
For BrainScaleS-2 systems, the clock signals are generated by a \gls{pll} developed by collaboration partners at TU Dresden \cite{hoeppner2013pll}, which is not depicted in the block-level schematic.

To achieve coherency with the continuously evolving accelerated neuromorphic core, the BrainScaleS-2 chips are connected to an \gls{fpga} via a low-jitter high-speed serial link.
The link is accessed on the \gls{fpga} via the \emph{playback executor} that consumes \emph{playback programs} which it fetches from local memory.
The playback programs contain instructions from a custom instruction set that facilitate the \emph{timed release} of actions like the injection of events or \gls{ocp} commands into the chip.
Simultaneously, data coming from the chip is tagged with timing information by the executor and stored in memory as an \emph{experiment trace} for analysis.
Playback programs can be either compiled locally on the FPGA by an on-board processor or transferred into local memory via Ethernet.

\section{Verification Methods}
\label{sec:ms-if}
In the following paragraphs we present methods and tool flows developed for the verification of the BrainScaleS\hyp{}2 mixed-signal \gls{asic}.

\subsection{RTL verification}
\label{sec:veri-rtl}

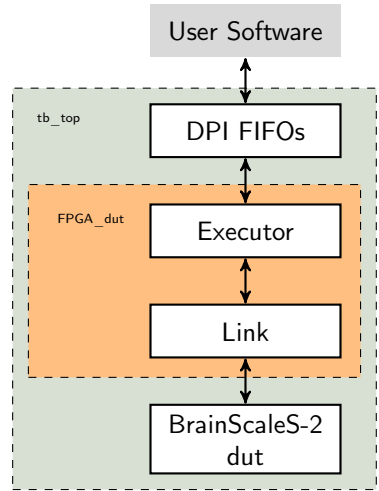
\begin{figure}
	\centering
	\pgfdeclarelayer{bg0}
	\pgfdeclarelayer{bg1}
	\pgfsetlayers{bg1,bg0,main}
	\begin{tikzpicture}[
						node distance=.6cm,
						thick,
						minimum width=2.5cm,
						minimum height=.7cm,
						>=stealth',
						font=\sffamily
						]
		\node[fill=gray!30] (user) {User Software};
		\node[draw, below=of user, fill=white] (dpi) {\gls{dpi} FIFOs};
		\draw[<->] (user) to (dpi);
		\node[draw, below=of dpi, fill=white] (pbexec) {Executor};
		\draw[<->] (dpi) to (pbexec);
		\node[draw, below=of pbexec, fill=white] (link) {Link};
		\draw[<->] (pbexec) to (link);
		\node[draw, below=of link, align=center, fill=white] (hx) {BrainScaleS-2\\dut};
		\draw[<->] (link) to (hx);
		\node[left=.2cm of pbexec.north west, anchor=north east, minimum width=0.1cm, minimum height=0.1cm] (hxfpga) {\tiny{FPGA\_dut}};
		\begin{pgfonlayer}{bg0}
			\node[draw, dashed, fit=(pbexec)(link)(hxfpga), inner sep=.25cm,fill=orange!50] (hxfpgabox) {};
		\end{pgfonlayer}
		\node[anchor=north west, minimum width=0.1cm, minimum height=0.1cm] (tbtop) at (hxfpgabox.north west|-dpi.north west) {\tiny{tb\_top}};
		\begin{pgfonlayer}{bg1}
			\node[draw, dashed, fit=(hxfpgabox)(hx)(dpi)(tbtop), inner sep=.2cm, fill=green!30] (tbtopbox) {};
		\end{pgfonlayer}
	\end{tikzpicture}
	\caption{Integration testbench for the BrainScaleS-2 system. The \gls{dut} are the BrainScaleS-2 design as well as parts of the \gls{fpga} responsible for playback program execution.}
	\label{fig:fpgastruc}
\end{figure}

The two important verification milestones are unit tests and integration tests.
Any design of sufficient complexity must employ both methods, as without unit tests it is unfeasible to localize bugs, while integration tests make sure that all interfaces are implemented correctly and there are no throughput mismatches.\cite{abrahams1998rtl}
The testbench for integration testing needs to encompass as much of the system as possible, ideally also including the majority of the user software stack.
Since the BrainScaleS System is controlled by an FPGA via playback programs which are generated by user software, it is convenient to use this interface for software-RTL co-simulation.
In the physical system, compiled playback programs are transported to the FPGA via Ethernet into local memory, from where they can be fetched and passed to the FPGA executor (cf.~\prettyref{sec:digital}).
In the simulation setup, we instantiate the BrainScaleS design together with the FPGA executor and their connecting link (cf.~\prettyref{fig:fpgastruc}).
The instances of analog macros like the \gls{pll} or \gls{sram} are replaced by behavioral models.
We then pass the compiled playback program via the \gls{dpi} \cite{sutherland2004integrating} into a blocking FIFO connecting to the FPGA executor which ensures the same execution pattern as in the physical system.
Errors are detected via software unit tests, as well as RTL assertions monitored by the simulator.
This setup is not only used for RTL verification, but also as a convenient reference for in-silico testing, since it is now possible to transparently execute a playback program in simulation or on the physical system and compare the results.

\subsection{Full-custom verification}
\label{sec:veri-fc}

Mixed-signal neuromorphic circuits as implemented in the BrainScaleS systems are designed to emulate complex biological mechanisms.
To allow for a flexible and faithful replication of the underlying models, the circuits must be tunable, which sometimes requires a large number of analog and digital parameters.
Both, the biological prototype as well as the neuromorphic replica can possess high-dimensional parameter spaces and a wide range of operating points.
Analog circuits are prone to parameter deviations due to mismatch effects and thus require additional calibration to reach a target operating point.
While individual components can often be unit-tested with conventional simulation strategies, assessability of a complete circuit's functionality is limited due to error propagation and inter\hyp{}dependencies of parameters.
Verifying such complex circuits is hence a challenge.

Software-driven simulation of such designs can aid the developer to increase pre-tape-out verification coverage, by allowing to programmatically generate stimuli and perform advanced analyses on recorded data.
Although inherently scriptable, the Cadence Virtuoso Analog Design Environment does not feature an ecosystem as rich as of more widely used programming languages \cite{cadence2019adexl,cadence2018ocean}.

\subsubsection{Interfacing analog simulations from Python}

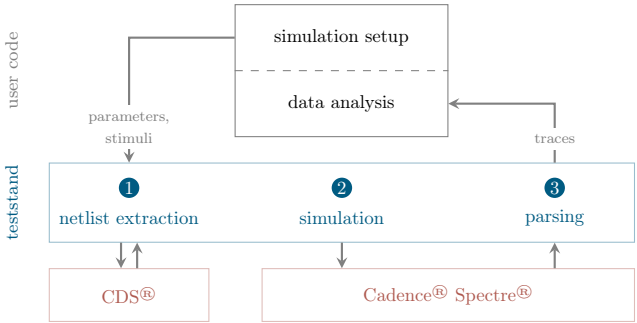
\begin{figure}
	\begin{center}
		\usetikzlibrary{shapes.misc,shapes,fit}
\usetikzlibrary{decorations.pathreplacing}
\usetikzlibrary{arrows.meta,decorations.pathreplacing,fadings,shapes,arrows}

\definecolor{orange}{HTML}{D7AA50}
\definecolor{purple}{HTML}{7A68A6}

\begin{tikzpicture}[scale=0.7, every node/.style={scale=0.7}]
	\node[gray,rotate=90,scale=1.2] at (-0.7,-1.25) {\footnotesize user code};
	\node[blue,rotate=90,scale=1.2] at (-0.7,-3.75) {\footnotesize teststand};
	
	\node[anchor=north,minimum width=4cm,minimum height=2.5cm,draw=gray] (u1) at (5.5,0) {};
	\node[yshift=+0.625cm,minimum width=4cm,align=center] (u11) at (u1)
		{simulation setup};
	\node[yshift=-0.625cm,minimum width=4cm,align=center] (u12) at (u1)
		{data analysis};
	\draw[dashed,gray] (u1.west) -- (u1.east);
	
	\node[below right,minimum width=3cm,minimum height=1.5cm] (t1) at (0,-3) {};
	\node[shape=circle,inner sep=1pt,fill=blue,yshift=0.8em] at (t1) {\color{white}1};
	\node[yshift=-0.8em] at (t1) {\color{blue}netlist extraction};
	
	\node[below right,minimum width=3cm,minimum height=1.5cm] (t2) at (4,-3) {};
	\node[shape=circle,inner sep=1pt,fill=blue,yshift=0.8em] at (t2) {\color{white}2};
	\node[yshift=-0.8em] at (t2) {\color{blue}simulation};
	
	\node[below right,minimum width=3cm,minimum height=1.5cm] (t3) at (8,-3) {};
	\node[shape=circle,inner sep=1pt,fill=blue,yshift=0.8em] at (t3) {\color{white}3};
	\node[yshift=-0.8em] at (t3) {\color{blue}parsing};
	
	\node[below right,minimum width=3cm,minimum height=1.0cm,draw=red!40] (c1) at (0,-5) {};
	\node[] at (c1) {\color{red}CDS\textsuperscript{®}};
	
	\node[below right,minimum width=7cm,minimum height=1.0cm,draw=red!40] (c2) at (4,-5) {};
	\node[] at (c2) {\color{red}Cadence\textsuperscript{®} Spectre\textsuperscript{®}};
	
	\node[below right,minimum width=11cm,minimum height=1.5cm,draw=blue!40] (t) at (0,-3) {};

	\draw[-stealth,thick,gray] (u11.west) -- (u11.west-|t1.north) -- (t1.north);
	\draw[stealth-,thick,gray] (u12.east) -- (u12.east-|t3.north) -- (t3.north);
	
	\draw[-stealth,thick,gray] ([xshift=-1ex]t1.south) -- ([xshift=-1ex]c1.north);
	\draw[-stealth,thick,gray] ([xshift=+1ex]c1.north) -- ([xshift=+1ex]t1.south);
	
	\draw[-stealth,thick,gray] (t2.south) -- (t2.south |- c2.north);
	\draw[stealth-,thick,gray] (t3.south) -- (t3.south |- c2.north);

	\node[anchor=south,fill=white,align=center] at ([yshift=+0.7em]t1.north) {\footnotesize\color{gray}parameters,\\\footnotesize\color{gray}stimuli};
	\node[anchor=south,fill=white,align=center] at ([yshift=+0.7em]t3.north) {\footnotesize\color{gray}traces};
\end{tikzpicture}
	\end{center}
	\caption{Structure of a teststand-based simulation including the interaction with the Cadence Design Suite.}
	\label{fig:schematic_teststand}
\end{figure}

\begin{figure*}
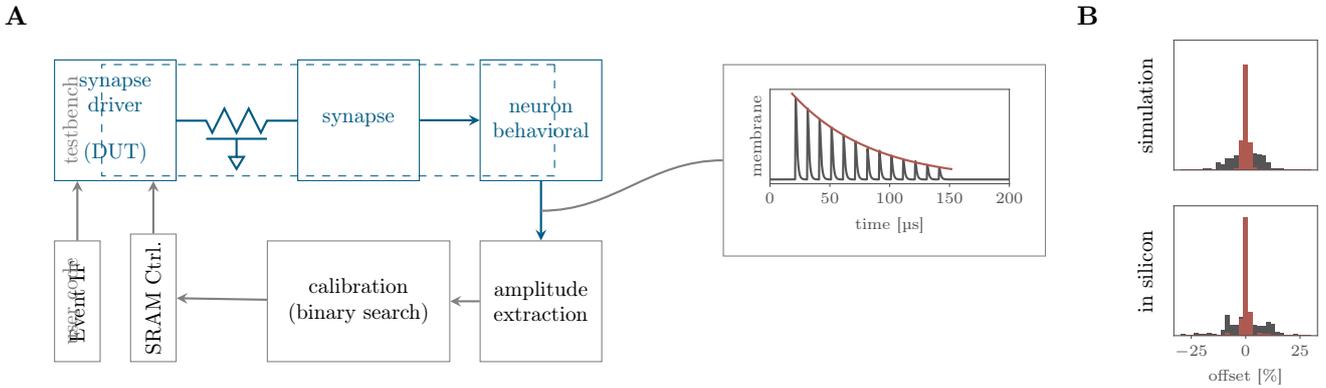

	\begin{center}
		\begin{tikzpicture}[]%
	\begin{scope}[shift={(-12.0,-1.0)},scale=0.8,transform shape]
		\node[anchor=south west,minimum width=2cm,minimum height=2cm,draw=blue,blue,align=center] (drv) at (0.0,2.0) {synapse\\ driver\\[1em] (DUT)};
		\node[anchor=south west,minimum width=2cm,minimum height=2cm,draw=blue,blue,align=center] (syn) at (4.0,2.0) {synapse};
		\node[anchor=south west,minimum width=2cm,minimum height=2cm,draw=blue,blue,align=center] (nrn) at (7.0,2.0) {neuron\\ behavioral};

		\draw[blue,thick] (2.0,3.0) -- ++(0.5,0) -- ++(0.1,-0.2) -- ++(0.2,0.4) -- ++(0.2,-0.4) -- ++(0.2,0.4)
			-- ++(0.2,-0.4) -- ++(0.1,0.2) -- ++(0.5,0);
		\draw[blue,thick] (2.0,3.0) ++(0.5,-0.3) -- ++(1.0,0.0);
		\draw[blue,thick] (2.0,3.0) ++(0.5,-0.3) ++(0.5,0.0) -- ++(0.0,-0.3)
			-- ++(-0.12,0.0) -- ++(0.12,-0.2) -- ++(0.12,0.2) -- ++(-0.12,0.0);

		\node[anchor=north west,minimum width=2cm,minimum height=0.75cm,draw=gray,align=center,rotate=90] (mct) at (1.25,-1.0) {SRAM Ctrl.};
		\node[anchor=north west,minimum width=2cm,minimum height=0.75cm,draw=gray,align=center,rotate=90] (pst) at (0.00,-1.0) {Event IF};
		\draw[gray,thick,-stealth] (mct) -- (mct |- drv.south);
		\draw[gray,thick,-stealth] (pst) -- (pst |- drv.south);
		
		\node[anchor=south west,minimum width=2cm,minimum height=2cm,draw=gray,align=center] (prc) at (7.0,-1.0) {amplitude\\ extraction};
		\node[anchor=south west,minimum width=3cm,minimum height=2cm,draw=gray,align=center] (cal) at (3.5,-1.0) {calibration\\ (binary search)};
		\draw[gray,thick,-stealth] (prc) -- (cal);

		\draw[blue,thick,-stealth] (syn) -- (nrn);
		\draw[blue,thick,-stealth] (nrn) -- (prc);
		\draw[gray,thick,-stealth] (cal) -- (mct);
		\coordinate (zoom) at ($(nrn)!0.5!(prc)$);
		
		\node[fit=(drv)(nrn),draw,blue,dashed] (tbc) {};
		\node[fit=(pst)(prc)] (usr) {};
		\node[gray,anchor=south west,rotate=90,above,yshift=0.2cm] at (tbc.west) {\vphantom{gl}testbench};
		\node[gray,anchor=south west,rotate=90,above,yshift=0.2cm] at (usr.west) {\vphantom{gl}user code};
	\end{scope}
	
	\begin{scope}[shift={(-1.6,0.0)},scale=0.8,transform shape]
		\node[anchor=south west,draw=gray] (trace) at (-2,-0.5) {\input{figures/teststand/example_stp/trace.pgf}};

		\node[anchor=north west] (hist_sim) at (5,3.5) {\input{figures/teststand/example_stp/hist_sim.pgf}};
		\node[below,rotate=90,xshift=-1.5cm,yshift=0.3cm] at (hist_sim.north west) {\vphantom{gl}simulation};
		
		\node[anchor=south west] (hist_hw) at (5,-3.0) {\input{figures/teststand/example_stp/hist_hw.pgf}};
		\node[below,rotate=90,xshift=-1.5cm,yshift=0.3cm] at (hist_hw.north west) {\vphantom{gl}in silicon};

		\draw[gray,thick] (zoom) to[out=0,in=180] (trace.west);
	\end{scope}

	\node at (-12.5,2.8) {\bfseries A};
	\node at (  1.6,2.8) {\bfseries B};
\end{tikzpicture}
	\end{center}
	\caption{Examplary \gls{mc} calibration workflow using teststand for an \gls{stp} circuit. \textbf{A} Testbench and corresponding program flow. \textbf{B} Offset distribution prior to (black) and after (red) calibration for a virtual as well as an in-silicon circuit array. In both cases the calibration was performed on 128 samples/instances.}
	\label{fig:teststand_calibration}
\end{figure*}

We implemented the Python module \emph{teststand} to provide tight integration of analog simulations into the language's ecosystem.
Teststand does not implement a new simulator but rather represents a thin layer to interface with the Cadence Spectre simulator and other tools from the Cadence Design Suite.

Netlists are directly extracted from the target cell view as available in the design library.
The data is accessed by querying the database via an OCEAN script executed as a child process.
Teststand then reads the netlist and modifies it according to the user's specification.
In addition to the schematic description, Spectre netlists also contain simulator instructions.
Teststand generates these statements and hence potentially supports all features provided by the backend.
Specifically, the user can define analyses to be performed by the simulator, such as DC, AC, and transient simulations.
\gls{mc} analyses are supported as well and play an important role in the verification strategies presented below.

The user specifies the simulation including e.g. stimuli, parameters, and nodes to be recorded using an object-oriented interface that resembles Spectre simulation instructions.
\begin{lstlisting}[style=pythonstyle]
cell = ('mylib', 'mycell', 'schematic')
nets = ['I0.mynet']

teststand = Teststand(cds_lib, cell)
tran = TransientAnalysis('tran', 1e-3)
simulation = Simulation(
	[tran], params, save=nets)
result = teststand.simulate(simulation)
\end{lstlisting}
The \texttt{simulate()}-call executes Spectre as a child process.
Basic parallelization features are natively provided via the \textit{multiprocessing} library.
Scheduling can be trivially extended to support custom compute environments.
The simulation log is parsed and potential error messages are presented to the user as Python exceptions.

Results are read and provided to the user as structured NumPy arrays.
This allows to resort to the vast amount of data processing libraries available in the Python ecosystem to process and evaluate recorded data.
Most notably, this includes NumPy \cite{oliphant2006guide}, SciPy \cite{Scipy2001}, and Matplotlib \cite{hunter2007matplotlib}.
As a side effect, the latter allows to directly generate rich publication-ready figures from analog circuit simulations.

\subsubsection{Monte Carlo calibration}

Teststand's benefits become most clearly visible in conjunction with \gls{mc} simulations, which allow to asses the performance of a circuit under the influence of random variations in the production process.
Traditionally, developers analyze a circuit's performance under statistical parameter variations that mimic in-silicon imperfections to allow them to anticipate post-tape-out functionality and yield.
Moreover, by fixing the \gls{mc} seed a set of \emph{virtual instances} can be obtained, which can be individually parameterized and analyzed, similar to an array of actual in-silicon instances of the design.

Such simulations can be iteratively and algorithmically modified.
This concept can be used to optimize bias and reference parameters $\theta_\text{hw}$ of a design to reach a desired operating point determined by a set of model parameters $\theta_\text{model}$.
In the case of a neuromorphic circuit these can for example be given by a set of potentials, conductances, and time constants.   %
Such a \gls{mc} calibration can be performed on each sample individually to also equilibrate mismatch-induced effects.

The approach to find a suitable parameter set generally depends on both model and circuit.
One possible strategy is based on iteratively reconfiguring and probing the design's behavior.
An effective implementation will likely be based on a binary search.
This method is particularly useful for parameters that are intended to be kept constant during operation, e.g. to compensate for a fixed offset.
In other scenarios it might be desirable to find and measure a transformation between the model's and circuit's parameter spaces $\theta_\text{hw}(\theta_\text{model})$ and make it persistent.
These data can then later be reused to perform one or multiple benchmarks on the calibrated instance, incorporating potentially different operating points.

These calibration algorithms are -- when required -- often implemented only after tape-out.
Already implementing them for simulated instances, however, brings several major advantages.
It allows the designer to determine a suitable calibration range and resolution and estimate the post-calibration yield.
The co\hyp{}development of circuits and algorithms leads to better hardware but also improved software, and might reveal details in their interplay otherwise potentially overlooked.
Especially for complex circuits with high-dimensional parameter spaces there might occur multidimensional dependencies which can be hard to resolve.
Actually calibrating such a circuit as a whole might reveal insufficient parametrization that would not have been found in tests of individual sub-components.
In order to uncover potential regressions due to modifications to a circuit, simulations based on teststand can easily be automated and allow continuous integration testing for full-custom designs.

For the BrainScaleS systems, the use of teststand has lead to large increase of in-silicon usability.
It was used throughout the verification of various components of the BrainScaleS-2 \glspl{asic}, including the current neuron implementation \cite{mueller2017phd,aamir2018adex}.
As a more compact example of teststand usage, we want to present a verification strategy for the BrainScaleS-2 synapse driver circuit, focussing on the analog implementation of \gls{stp}.
The testbench shown in \prettyref{fig:teststand_calibration} is centered around the synapse driver as the design under testing.
The latter is accompanied by an instance of the synapse circuit.
To mock parasitic effects due to the synapse array's spatial extents, an RC wire model based on post-layout extractions is inserted in between the two instances.
Finally, a simple neuron circuit based on ideal components is included in the testbench, integrating the post-synaptic currents to form the characteristic post-synaptic potentials.

The testbench is controlled from Python code using teststand.
Both input interfaces, the \gls{sram} controller as well as the event interface receiver, are mocked in Python, allowing for the verification of the entire design in a realistic scenario, beginning with accessing the configuration memory and then moving to processing of synaptic events.
The synapse driver is exposed to predefined input spike trains consisting of a series of equidistant events.
The design's response is recorded and then processed using tools from the Python ecosystem in order to extract parameters from the biological model \cite{tsodyks97neural}.
Thus, quantities as the circuit's synaptic utilization and the recovery time constant, describing the decay and re-uptake of synaptic resources, can be benchmarked against specification and constraints.
More importantly, a mismatch-induced offset in synaptic efficacy can be extracted and compared across multiple virtual synapse drivers.
Following a binary search based on their deviation from a target value, a \SI{4}{\bit} offset calibration parameter in the \gls{dut}'s configuration memory is iteratively reprogrammed, minimizing the offset.
Implementing this calibration routine before tape-out allowed to fully judge the circuits usability.
\prettyref{fig:teststand_calibration} includes histograms of the extracted offsets for 128 synapse driver instances, prior to and after calibration.
Applying the same calibration methodology to the taped-out circuits resulted in very similar distributions. 
While certainly relying on the quality of the models provided with the process design kit, these results show that the advanced verification methods facilitated by teststand allow to successfully pre-asses the behavior of even complex full-custom circuit designs that require calibration.

\section{Physical Implementation}
\label{sec:phys-impl}
\begin{figure*}[htbp!]
	\includegraphics[width=\textwidth]{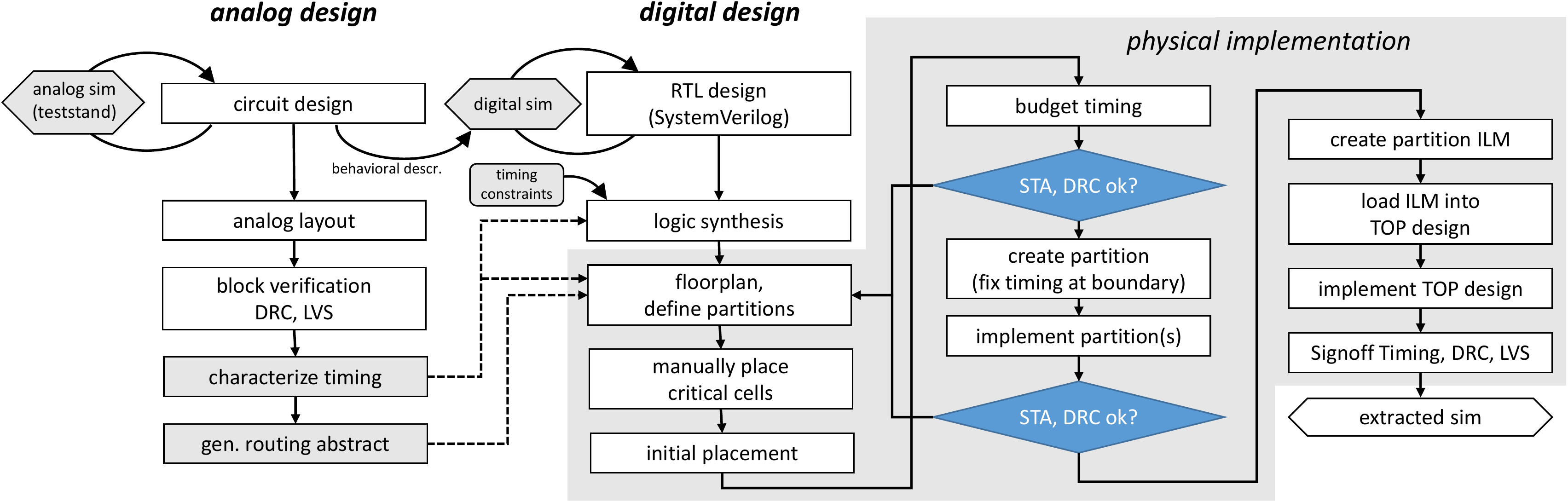}
	\caption{Simplified mixed-signal design flow. Shaded elements are explained in more detail in the main text. In digital physical design (large grey block), only non-standard or noteworthy steps are explcitly mentioned. The \textit{implement partition(s)} and \textit{implement TOP design} steps each comprise a complete place-and-route implementation flow, including in-place timing optimizations, clock tree synthesis, and \gls{sta}.}
	\label{fig:inn-dsgnflw}       %
\end{figure*}

Physical implementation describes the process of generating an \gls{asic} layout from a netlist description.
It is part of a usually customized design flow which is applied during the overall design process.
For BrainScaleS-2 we apply separate flows for analog and digital design as illustrated in \prettyref{fig:inn-dsgnflw}.
Analog layout is carried out using Cadence Virtuoso and shall not be covered in this paper.

We are using a digital top-level description for all BrainScaleS chips (cf. \prettyref{sec:digital}), thus top-level chip assembly is carried out in the digital design flow, using Cadence Innovus.
Depending on the complexity of the specific design, we follow a hierarchical implementation approach using separate design partitions which might be instantiated multiple times in the design.
Besides this re-usability, partitioning the design has the main advantage of a dedicated implementation approach per partition, for example optimized for a purely digital partition, or a partition having or containing a mixed-signal interface.

Our digital logic is written in SystemVerilog, and partially in VHDL. %
The gate-level netlist which is the basis for physical implementation is generated during logic synthesis, where the RTL description of the logic is mapped to a standard cell library \cite{tsmc2010stdcell}. %
Blocks with more complex functionality (such as large \gls{sram} blocks, \glspl{pll}) need to be provided as pin-level macros and are directly instantiated already in the RTL description.
Both, logic synthesis and physical implementation require a pin-level characterization of the signal timing of the blocks in order to correctly analyze static timing of the whole design.
Characterization is also required for our analog neuromorphic circuits, which are directly instantiated in the SystemVerilog source.
The following section covers methods that we have developed for this purpose.

\subsection{Timing analysis at mixed-signal interfaces}
\label{sec:ms-timinchar}

The \gls{ppu} has local memory for program data and vector operations, but also accesses the memory that has been implemented into the synapse array for digital synaptic weight storage.
Thus, the synapse array can be accessed row-wise by the \gls{ppu}, with every column of the array being directly connected to the synapse memory access controller.
The resulting data bus has a width of$\SI{8}{\bit}$ times number of synapses per row.

It is desirable to minimize the access time to the synapse weight storage in order to maximize the weight update rate during plasticity operation \cite{friedmann13phd}.
To facilitate this and reduce access latencies, a full-custom SRAM controller has been implemented in the synapse array \cite{hock14phd}.
It has a fully synchronous digital interface towards the PPU that is designed to operate at the maximum targeted PPU operating frequency of $\SI{500}{\mega\hertz}$.
All circuits behind the registers of this interface are covered by the verification steps in the analog design flow and do not have to be taken into account for timing analysis at the interface.
As a consequence, only the communication with the interface registers needs to be verified in order to ensure correct functionality at this mixed-signal interface.

\subsubsection{Timing characterization of anncore}
\label{sec:anncore_timing_char}

Synchronous digital circuits are commonly implemented using a set of standard cells that implement logic gates and memory elements (e.g. flip-flops).
In contrast to analog circuits, performance of digital circuits is not evaluated by transistor-level simulations, but by \gls{sta} \cite{bhasker2009_sta} which verifies whether setup and hold timing constraints are met for all flip-flops in the design, thus, whether the design is able to operate at a given clock period. 
\gls{sta} requires information about setup, hold, and clock-to-output time of flip-flops, delays through logic gates, as well es external capacitive load on cells and the propagation delay on signal wires.
Among those, all cell-related delays are dependent on the actual wiring of the cell, operating conditions and process corner.
Therefore, not a single value can be given for e.g. a gate propagation delay, but the cells rather have to be characterized for several sets of conditions, usually covering typical values that arise during operation.
Results are stored in a timing library file containing either look-up table data or a current source model \cite{bhasker2009_sta}.
For each combination of process corner and operating conditions that should be analyzed one such library is provided by the standard cell vendor.
When calculating \gls{sta}, the tools are allowed to extrapolate from and interpolate in between the given values.

Commercial tools exist for characterizing custom designed standard cell libraries, as for example Cadence Liberate.
These tools can automatically determine the relevant signal paths through circuits representing logic gates or flip-flops and then carry out a series of analog simulations in order to determine the aforementioned delay values under a certain set of conditions.
However, these tools are scarcely configurable for automatically analyzing complex VLSI circuits, like the described synapse memory interface of the anncore.
For this reason, a Python-based characterization framework has been developed in \cite{hartel2016phd}.

Sequential input pins with a timing relation to a clock input are characterized for capacitance, setup and hold time. Output pins associated to a clock input are characterized for clock-to-output delay and load\hyp{}dependent output transition time.
For example all data pins of the synapse memory interface belong to this category.
Non-sequential input pins are solely characterized for their capacitance and output pins for their transition time.
Potential timing requirements on these pins need to be defined externally.
This includes for example pins of custom \gls{sram} arrays, static control pins, and the event interface.

The clock signal of the synapse array memory interface's registers is distributed in a fly-by manner (see \prettyref{fig:mx-synraminterface}), along the edge of the synapse array.
This edge has a length of $\SI{1.5}{\milli\meter}$ in the current BrainScaleS-2 chip.
Since no balanced clock tree exists for these registers, a correct characterization of the resulting spread in timing constraints is one of the most crucial results of this characterization.

The digital timing of the anncore is characterized after completion of the analog design process and the resulting data is stored in a timing library file \cite{bhasker2009_sta}.
It can then directly be instantiated in the RTL code and is treated as a macro block throughout the digital design flow.
For the layout of the current anncore (see \prettyref{sec:anncore_abstract}), a spread in setup and hold times of approximately $\SI{150}{\pico\second}$ has been determined.
Most notably, the setup-and-hold window of the data pins which usually lies \textit{around} the clock edge of a flip-flop lies up to $\SI{600}{\pico\second}$ \textit{after} the related edge at the clock pin, due to the internal delay on the clock signal.

\bigbreak
For the digital design implementation of the current BrainScaleS-2 \gls{asic}, we have used a standard bottom-up hierarchical synthesis flow with Synopsys DesignCompiler to obtain a single shared implementation for the two \gls{ppu} instances.
As a first step during subsequent physical implementation the floorplan needs to be laid out.
The illustrated floorplan of the current BrainScaleS-2 full-size \gls{asic} is depicted in \prettyref{fig:hx_fp}.
Non-standard floorplanning and further physical implementation steps will be described in the following subsections.

\subsection{Anncore abstract view}
\label{sec:anncore_abstract}

\begin{figure}
	\includegraphics[width=\columnwidth]{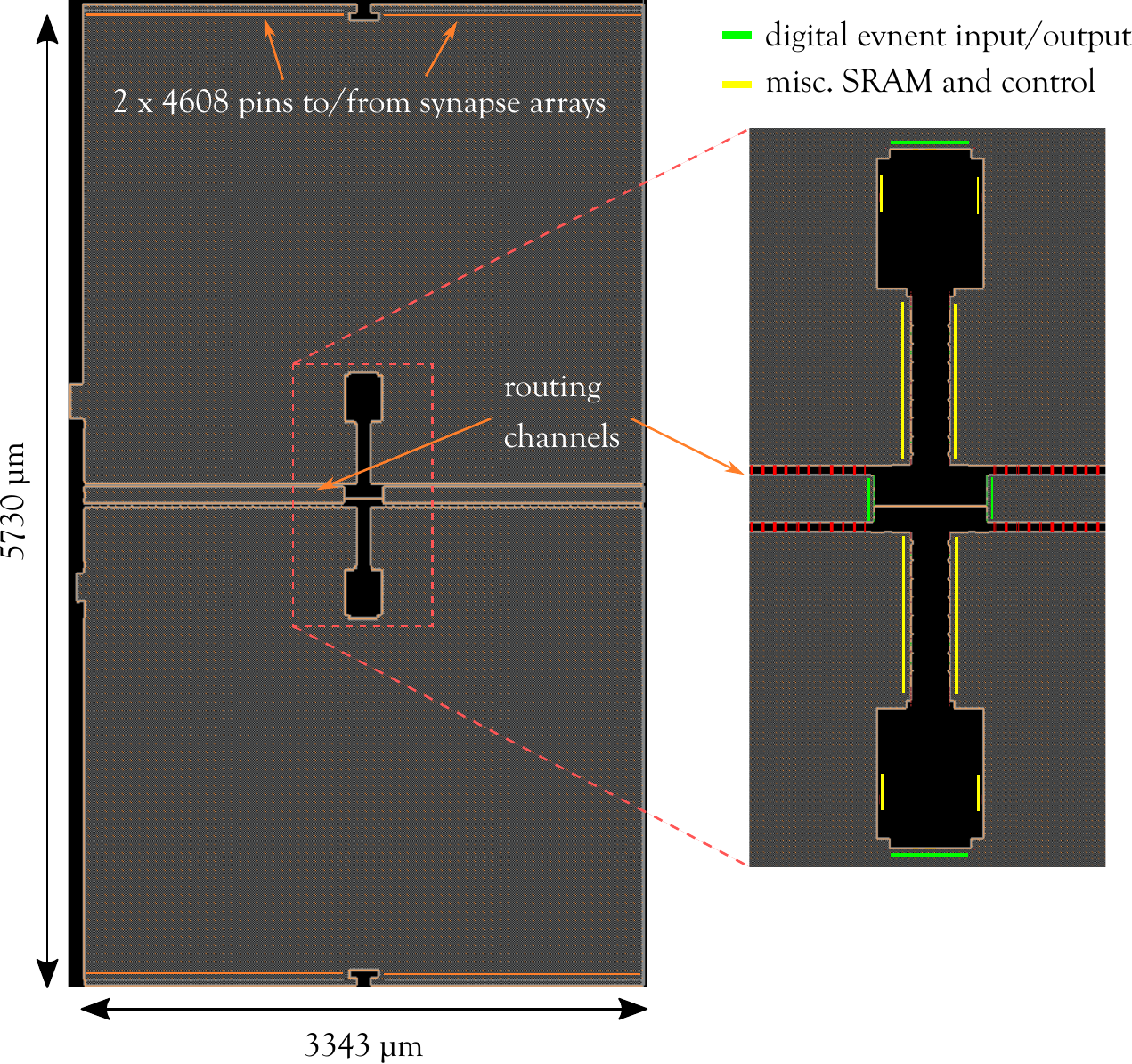}
	\caption{Abstract view of the anncore. Zoom-out: center cut-out with digital configuration pins of neuron circuits, capacitive memory and PADI bus.}
	\label{fig:annc_abstract}       %
\bigbreak
	\includegraphics[width=\columnwidth]{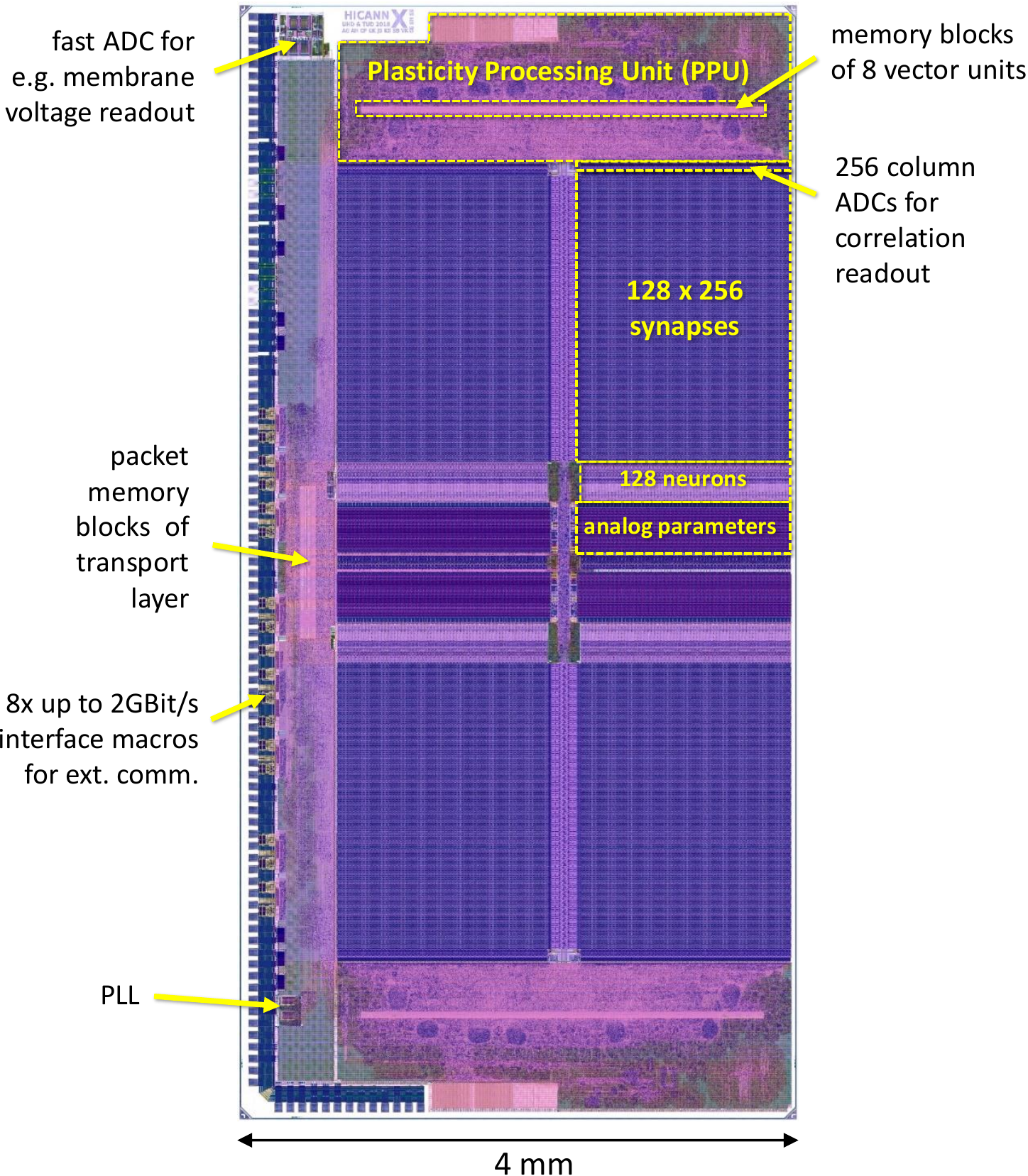}
	\caption{Layout of the current BrainScaleS-2 full-size \gls{asic}. It contains 512 neuron circuits and $\num{131072}$ synapse circuits which are arranged in 4 quadrants. Data lines of the synapse arrays and the column \glspl{adc} are directly connected to the \glspl{ppu} at the top and bottom edges. Each \gls{ppu} contains 8 vector units with dedicated memory blocks in addition to the general-purpose processor part. Analog quantities like membrane voltages or outputs of the analog parameter storage can be digitized on-chip by the \gls{adc} in the top left edge. Its data is merged with control and neural event data in the digital control part (left edge). Data transmission is secured by a custom developed transport layer, the physical interface consists of 8 SerDes blocks \cite{scholze2012link} with a data rate of up to $\SI{2}{\giga\bit\per\second}$ per block.}
	\label{fig:hx_fp}       %
\end{figure}

All analog circuits of the BrainScaleS architecture are arranged such that they are combined into one large analog macro block (anncore).
When implementing full-sized ASICs with up to 512 neurons and 256 synapse circuits per neuron, we split the resulting \gls{vlsi} synapse array into several subunits because resistance and capacitance of the long wires would lead to undesirably high energy consumption and internal signal delay.
The drawback is an increased number of pins that result at the split edges and require additional routing when connecting with the digital logic.
In case of the current BrainScaleS-2 chip we considered a 4 quadrant layout as a good compromise between energy consumption and routability.
It is illustrated in \prettyref{fig:hx_fp}, where the top right quadrant is illustrated.
Two halves of the array are arranged such that the neuron circuits are located at the horizontal symmetry axis in order to minimize vertical wire capacitances.
To balance horizontal wire capacitances and routability we choose to introduce a vertical split which adds additional row drivers and according pins at the split edges.
The quadrants have been arranged in a way that all control pins are facing towards a cut-out in the center of anncore (see zoom-out in \prettyref{fig:annc_abstract}).
The strategies we developed to connect to the pins in this center cut-out will be described in the following.

During physical implementation, abstracted layout data are required to floorplan the design and connect the block using the auto-router.
We generate these using Cadence Abstract Generator, with a few non\hyp{}standard tweaks to obtain a routable block, since physical size, shape, and number of pins pose several challenges to the standard abstract generation. 

\begin{figure*}[t]
	\begin{center}
		\NewDocumentCommand{\busref}{som}{\texttt{%
		#3%
		\IfValueTF{#2}{[#2]}{}%
		\IfBooleanTF{#1}{\#}{}%
		}}

		\begin{tikzpicture}
			\node at(0.0,0.2) {\textbf{A}};
			\node[anchor=north west] at (0,-0.3) {
				\begin{tikztimingtable}[%
						timing/.style={},
						timing/slope=0.1,
						x=2.3ex,
						timing/rowdist=4ex,
						timing/name/.style={font=\sffamily\scriptsize},
						anchor=center
					]

					\busref{select[4:0]} 	& 3.9D{n-1} 0.2U 7.8D{n} 0.2U 7.9D{n+1} \\
					\busref{address[5:0]}	& 7.9D{n-1} 0.2U 7.8D{n} 0.2U 3.9D{n+1} \\
					\busref{pre}		& 7.9L 0.2U 3.6H 0.2U N(BA1) 8.1L \\
					\busref{stable}		& 9.9L 0.2U 5.8H 0.2U 3.9L \\
					\busref{pulse}		& 12L N(CA1) 6H 2L \\
					\begin{extracode}
						\begin{background}
							\foreach \n in {-8,-4,...,12}
							\draw[dotted] (\n + 8,-\nrows*2-1) -- (\n + 8,2) node[above] {\scriptsize\SI{\n}{\nano\second}};
						\end{background}
						\draw[black,-stealth](BA1|-row3.mid)to[out=0,in=180](CA1|-row5.mid);
					\end{extracode}
				\end{tikztimingtable}
			};

			\node at(10.0,0.2) {\textbf{B}};
			\node[anchor=north west] at (10,0) {
				\input{figures/backend/source/ei_slack_histo.pgf}
				};
		\end{tikzpicture}
	\end{center}

	\caption{\textbf{A} The event interface consists of the row select and event addresses as well as three timing signals for the synapse driver. \textbf{B} Slack distribution w.r.t. pulse pins at the event interface.}
	\label{fig:padi_timing}
\end{figure*}

The anncore abstract is illustrated in \prettyref{fig:annc_abstract}.
Approximately $85\,\%$ of the pins are made up by the interfaces for synapse memory access and column ADC readout.
These pins are placed at the top and bottom edges of the anncore to facilitate direct access by the adjacent \glspl{ppu}.
The other $15\,\%$ of the pins consist of \gls{sram} and auxiliary control pins for the neuron configuration, the capacitive memory and the event communication. %
They are placed at the row-ends of their connected circuits for neurons and capacitive memory, and at the bottom edge of the event interface columns, all facing towards the cut-out in the center.

All control logic, including power supply, the according clock tree and the interface to the top-level control need to be placed in this cut-out area.
It has a size of approximately $\num{1440} \times \num{225} {\mathrm \mu m}^2$ with an area of approximately $\SI{2e5}{\square\micro\meter}$ being available for standard cell placement due to the dumbbell-shaped outline.
This would allow for roughly 30\,k flip-flops of minimum size at 100\,\% placement density. %
All but the two topmost metal layers are available for routing; the two topmost layers are exclusively used for power distribution.
Pins to the analog circuits are spread over the complete boundary, while care has been taken to optimize accessibility by the auto-router: they have been placed on layers with horizontal/vertical preferred routing direction depending on the edge, and blockage generation around the pins has been optimized for routability, per layer.

Access to this area has been enabled by means of two routing channels that have been left open during analog layout.
Three horizontal routing layers are available inside these channels and they have been sized in a way to accommodate routing of all required interface signals.
The generation of placement blockage in the generated abstract view has carefully been tuned to represent the actual outline of only the layout of metal layers defining the cut-out and channels, and not the covering power distribution layers (cf. \prettyref{fig:annc_abstract}).

Standard cells can therefore be placed inside the cut-out and the channels but the tool is restricted to only place buffers within the channels that are required to meet the timing constraints.
This allows keeping the routing channels as small as possible while still being able to achieve timing closure.
Bus guides have been used to guide the auto-router and use one channel for inbound and one channel for outbound signals, only.
All corresponding interface logic has been constrained to be placed in the proximity of the channel entry areas.

\subsection{Mixed-signal event input}
\label{sec:padi-impl}

Neural events are injected into the synapse drivers using four event interface buses in each half of anncore.
Each bus consists of four signals \texttt{address[5:0]}, \texttt{select[4:0]}, \texttt{pulse}, and \texttt{stable}.
These signals are generated by flip-flops in the event handling logic and are required to keep below a maximum skew of $\SI{200}{\pico\second}$ at the according anncore pins (see undefined regions in \prettyref{fig:padi_timing} A).
Since the inputs to the anncore have no synchronous relation to a clock signal, the timing to these pins cannot be constrained by a sequential relation, like the standard setup and hold conditions between two flip-flops.
The signals rather have to be treated like a source-synchronous bus with a strobe signal as a reference signal and all bus signals must be constrained to stay within a maximum skew compared to the strobe signal.
From a functional point of view, the \texttt{pulse} signal acts as this strobe signal (cf. \prettyref{sec:anncore_syndrv} and \prettyref{fig:padi_timing} A).
While allowing for a clock skew of $\SI{50}{\pico\second}$ to the registers generating these signals, they have been constrained for a maximum skew of $\SI{150}{\pico\second}$ w.r.t. the \texttt{pulse} signal using the following constraints:

\lstset{style=tclstyle}
\begin{lstlisting}
for {set i 0} {$i < 8} {incr i} {
  set ei_signals($i) <collect address, select, stable signals>
}
for {set i 0} {$i < 8} {incr i} {
  foreach_in_collection consPin $ei_signals($i) {
    set_data_check -from pulse[$i] -to $consPin -setup -0.15
    set_data_check -from $consPin -to pulse[$i] -setup -0.15
  }
}
\end{lstlisting}

The mutual definition of a negative setup time between the signals results in a temporal window within which the signals must arrive at the anncore pins.
The above statements are part of the timing constraints which are used as an input already for synthesis.
They are interpreted equally to a regular setup constraint and the tool fixes violations during setup-time optimization steps.
The resulting delay distribution of all affected signals is shown in \prettyref{fig:padi_timing} B.
In the typical and fast corner the delay values cover a range of $\SI{125}{\pico\second}$ and $\SI{75}{\pico\second}$, respectively.
In the slow corner, the spread is about $\SI{190}{\pico\second}$ which is perfectly within specification.

\begin{figure}
	\includegraphics[width=\columnwidth]{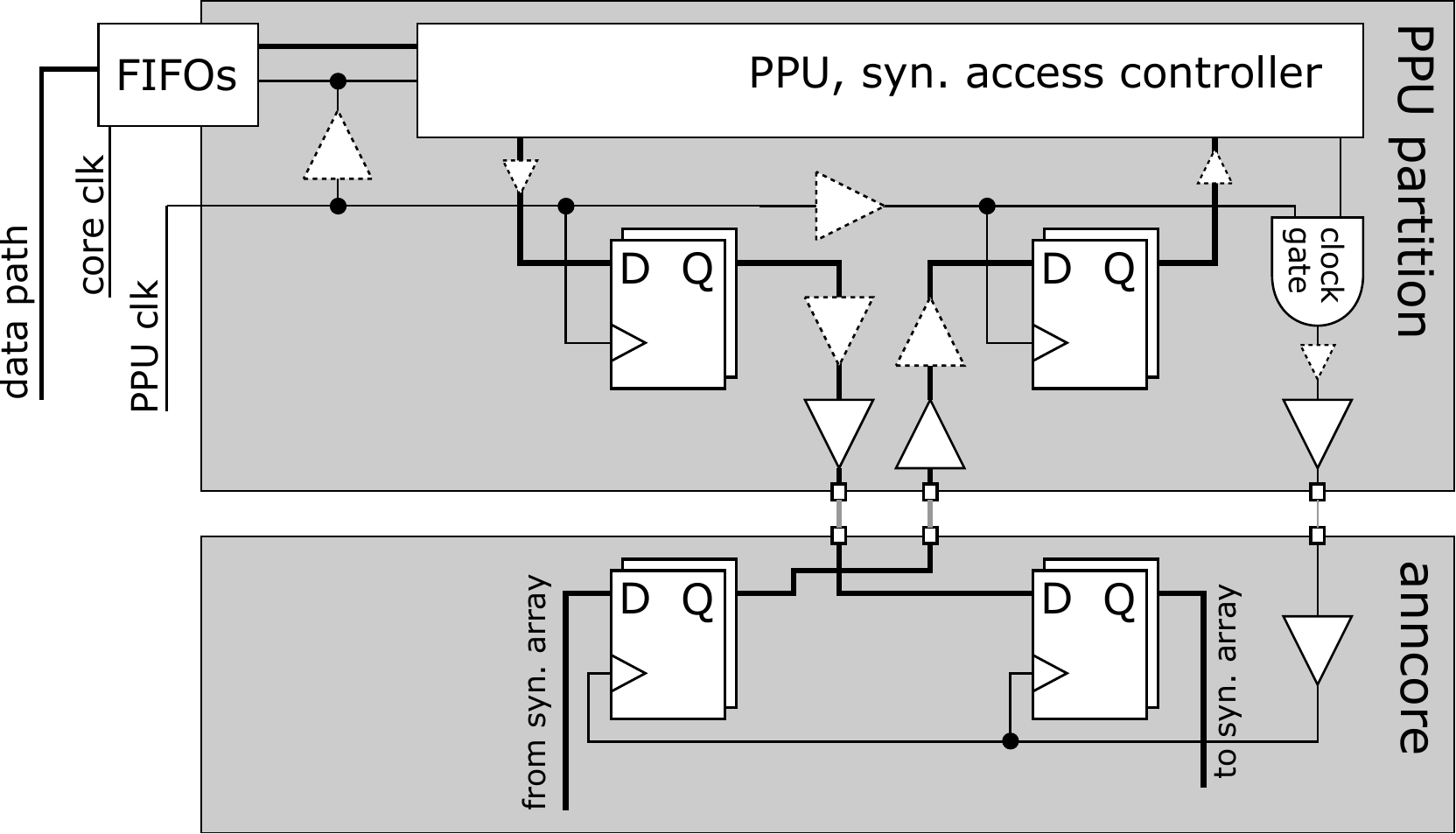}
	\caption{Data path and alock distribution network between the synchronous synapse memory interface and the corresponding logic in the plasticity processing unit. Grey wires illustrate direct connections between partition and anncore pins. Dashed buffer symbols denote signal and clock paths that can be timing optimized. All other routing and the placement of the solid buffers is fixed.}
	\label{fig:mx-synraminterface}       %
\end{figure}

\subsection{Partition interface timing}
\label{sec:before_placement}

In \prettyref{fig:hx_fp} the floorplan of the most recent version Brain\hyp{}ScaleS-2 chip is shown.
The two \glspl{ppu} are placed at the top and bottom edges as a copy of one implemented design partition.
Each \gls{ppu} has a purely digital interface to the digital control logic at its left edge and an interface to the anncore which is connected to anncore pins (see also \prettyref{fig:mx-synraminterface}).
Registers inside the PPU partition are connected to registers in the anncore while both receive the same clock.
This clock can be switched off towards the synapse array by a clock gate, controlled by the \gls{ppu}, to save dynamic power.
The main problem that comes with this configuration is the fact that this gated clock is yet to be implemented inside the PPU partition, thus has an initially unknown clock tree propagation delay.
Therefore the standard methods to derive the interface timing for the partition implementation are not applicable here.
We use the following approach to solve this, which could be considered a generic solution to such configurations:

In general, \textit{timing budgeting} using \textit{virtual in-place optimization} (IPO) provided by the Innovus tool is used to derive the partition interface timing before splitting off the \gls{ppu} design for separate implementation.
A preliminary place-and-route step and a provisional timing optimization is automatically run in this step to estimate the signal timing at partition boundaries.
During budgeting, a certain amount of available time on signal paths between two flip-flops before and after a partition boundary (\textit{slack}) is distributed between both sides, depending on provisional optimization results.
The changes made during optimization are then reverted and actual timing optimization has to be carried out during partition and top-level implementation, using the slack values that have been distributed to the respective signal pins.
However, since the involved algorithms assume that during later implementation steps the optimization engine will operate on both sides of the partition boundary, this cannot be applied to the signals directly connecting to the anncore since no buffers can be added outside the partition.
This affects all grey interconnect lines and routing inside the anncore (cf. \prettyref{fig:mx-synraminterface}).

To solve this problem the following method is applied for budgeting of the partitions' timing constraints before partitioning the design:
Pin locations at the partition boundary are fixed, adjacent to their anncore counterparts, in order to have predictable routing lengths between PPU partition and anncore.
Sufficiently sized buffers are constrained to be placed close to those pins inside the PPU partition, in order to fix the capacitive load on input pins and the drive strength at output pins, respectively (solidly drawn buffers in \prettyref{fig:mx-synraminterface}).
These buffer cells are already instantiated in the RTL description; they are merely up- or downsized in this step, according to their actual load.
After completion of these steps, a preliminary routing and \gls{sta} is run in order to determine the signal delays between partition boundary and anncore (grey interconnect lines and in-anncore routing are fixed at this step).
The result is then used as a fixed slack outside the partition, while the remaining slack is available for the timing paths inside the partition.

The delay between partition and anncore pins is determined in a similar fashion for the according clock signals.
In order to get the delay reported correctly, a clock buffer is placed and fixed close to the partition boundary, serving as a start point for the path segment between PPU partition and anncore.
The determined delay is then given to the clock tree synthesizer and is accounted as an external additional delay during clock tree synthesis.
Toghether with the strategy for fixing external signal delay, the setup condition for the anncore registers can be written as
\begin{equation}
(t_{\mathrm cp} + \Delta t_{\mathrm cp}) + t_{\mathrm dp} + \underbrace{t_{\mathrm dt} + t_{\mathrm co} + t_{\mathrm sut}}_{\mathrm delay~fixed} 
 \leq t_{\mathrm cp} + \underbrace{t_{\mathrm ct} + t_{\mathrm per}}_{\mathrm delay~fixed}\,,
\label{eq:sa_clk_su_cond}
\end{equation}
with $t_{\mathrm cp}$ being the clock tree delay inside the PPU and $\Delta t_{\mathrm cp}$ the skew after clock tree synthesis, $t_{\mathrm dp}$ the signal path delay inside the \gls{ppu} (logic and wires), $t_{\mathrm dt}$ the external signal delay between \gls{ppu} and anncore, $t_{\mathrm co}$ the clock-to-output time of the flip-flops inside the PPU, $t_{\mathrm sut}$ the setup time of the anncore register, $t_{\mathrm ct}$ the portion of the clock tree delay between PPU and anncore, and $t_{\mathrm per}$ the clock period.
This condition must be met by the tool during optimizations in the PPU partition.
To achieve this, the clock port to the synapse array and the related registers inside the \gls{ppu} partition are constrained into a separate \textit{skew group} which can subsequently be optimized separately by the clock tree synthesizer.
The maximum allowed skew within this group is set to $\SI{0}{\pico\second}$ to force the clock tree synthesizer to achieve as identical as possible $t_{\mathrm cp}$ at all those endpoints.
It is allowed to skew all other registers, if useful for timing optimization.

Ideally, the described setup condition could then be met by timing optimization steps during partition implementation.
However, zero skew cannot be realized by the clock tree synthesizer, especially over large spatial distances, as is the case along the anncore edges.
As a consequence, the resulting clock skew $\Delta t_{\mathrm cp}$ at the constrained registers has to be checked after clock tree synthesis is finished.
This maximum skew value is a timing uncertainty that could not be taken into account during calculation of the partition timing budgets.
Therefore, it has to be accounted for in the signal paths between \gls{ppu} partition and anncore by adding the skew value as a \textit{slack adjustment} to these paths in the scripts that are used for partition implementation.
This way, setup timing gets slightly overconstrained for most paths, yet we found no other way to safely account for the inevitable clock skew.
At least one iteration of the partition implementation design flow is necessary to obtain the skew values and add them to the scripts, ideally already before initial placement, to have consistent constraints throughout the design flow.

\subsection{Partition and top-level implementation}
\label{sec:rem_impl}

In the \gls{ppu} partition, each slice of the vector unit is connected to 32 synapse columns each and operates only on data local to the slice.
This spatial correlation results in a predictable implementation quality of the vector units themselves in terms of area and timing.
However, the vector control unit requires access to all synapse data, and the state values of the vector units.
Therefore, the critical path inside the PPU partition runs between registers in the outermost vector units \textit{through} the vector control unit which is located in the partition's geometrical center to the outermost synapse array data pins.
Since the responsible RTL designer left the group prior to tapeout we could not improve this path by e.g.~adding pipeline registers, for this chip revision.
\gls{ppu} partition implementation is carried out  using a standard physical design flow, including pre- and post-route in-place timing optimizations and the previously discussed modifications to clock tree synthesis and the slack adjustment.
Maximum expected clock frequency of the \gls{ppu} in the worst process corner is $\SI{245}{\mega\hertz}$, due to the aforementioned critical path.
Initial measurements, running a memory test on the full synapse array which was executed on the \gls{ppu} using the access path through the vector units yielded a maximum clock frequency of $\SI{400}{\mega\hertz}$.

The top-level implementation essentially follows a standard physical design flow as well, with two exceptions:
First, drivers to full-custom SRAM bitlines of the various configuration memories in the anncore center are placed close to their corresponding pins, to obtain equal parasitic load on those lines.
This is done automatically by means of a script that determines pin location and the connected cell, and places the cell at the closest legal location to the pin.
Second, the clock tree generation to the center cut-out in the anncore is constrained in a way to optimize balancing between flip-flops that are located outside and inside the cut-out area.
This is beneficial in terms of overall clock tree depth, thus power consumption on the clock tree, because balancing the tree globally would require an adaption of all clock sinks to the additional delay introduced by the routing channels into the anncore center.
A similar approach to the technique described in \prettyref{sec:before_placement} is taken to achieve good balancing:
all registers that are to be placed inside the anncore center are constrained into a dedicated skew group which is disjunct from the remainder logic, with no skew constraint set.
This way, the clock tree synthesizer can optimally balance inside and outside skew with respect to the anncore center.

As an implementation result, a total of $\num{33003}$ standard cells have been placed in the anncore center, at an average placement density of roughly 75\,\% in the $\SI{2e5}{\square\micro\meter}$ area and no issues in routability (see \prettyref{sec:anncore_abstract} for an area calculation).
Timing could be closed in all process corners, the target clock frequencies of $\SI{250}{\mega\hertz}$ and $\SI{125}{\mega\hertz}$ for link/event handling clock and on-chip bus clock, respectively have been met and proven in silicon (cf. \prettyref{sec:veri-fc}).

\section{Applications}
\label{sec:applications}
\begin{figure}
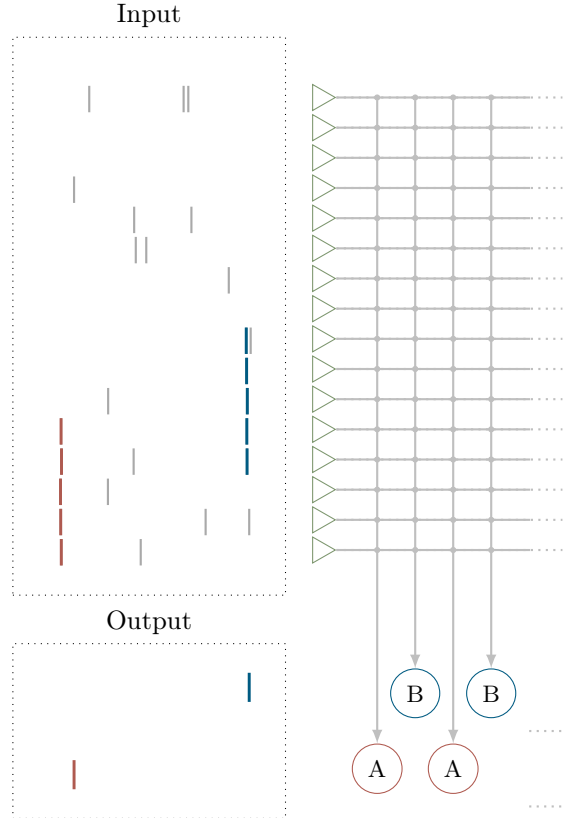

	\begin{tikzpicture}
	\pgfmathsetmacro{\N}{15}  %
	\node[draw,dotted,label=Input] (spikes_in) at (0, 0) {\input{figures/rstdp/input.pgf}};

	\node[draw,dotted,label=Output] (spikes_out) at (0, -5.5) {\input{figures/rstdp/output.pgf}};

	\foreach \i in {0,...,3}{
	       \ifodd\i{
		       \node[circle,draw,blue,text=black,font=\small] (output\i) at (3.0 + 0.5*\i, -5.0) {B};
	       }\else{
		       \node[circle,draw,red,text=black,font=\small] (output\i) at (3.0 + 0.5*\i, -6.0) {A};
	       }\fi
	}

	\foreach \i in {0,...,\N}{
		\node[regular polygon,regular polygon sides=3,draw,green,inner sep=2pt,shape border rotate=-90] (input\i) at (2.25, -3.1+\i*0.4) {};
	}

	\foreach \i in {0,...,\N}{
		\foreach \j in {0,...,3}{
			\node[circle,draw,lightgray,fill,scale=0.2] (synapse\i\j) at(3.0 + 0.5*\j, -3.1+\i*0.4) {};
		}
	}

	\foreach \i in {0,...,\N}{
		\node (virtual\i) at(3.5 + 3.0, -3.1+\i*0.4) {};
	}

	\foreach \i in {0,...,\N}{
		\draw[lightgray,thick] (input\i) edge ($(input\i)!0.65!(virtual\i)$) edge [dotted, thick] ($(input\i)!0.75!(virtual\i)$);
	}
	
	\foreach \i in {0,...,3}{
		\draw[-latex,lightgray,thick] (synapse\N\i) to (output\i);
	}

	\foreach \i in {0,1}{
		\draw[dotted,thick,lightgray] (5.0, -5.5-\i) -- (5.5, -5.5-\i);
	}
	
\end{tikzpicture}
	\caption{%
		Schematic illustration of the R-STDP experiment.
		The input consists of Poissonian background spikes in which two input patterns are embedded.
		The spikes of each source are sent to a single synapse driver (green triangles) to enter the synapse array.
		Even neurons (red) are trained to fire if the network is stimulated with pattern A, whereas the odd ones (blue) should fire if pattern B is applied.
		}
	\label{fig:overview_rstd}
\end{figure}

The BrainScaleS systems have been used for a wide range of experiments.
We have demonstrated porting of deep artificial neural networks to the wafer\hyp{}scale BrainScaleS\hyp{}1 system with in-the-loop training \cite{schmitt2017itl}.
The platform was also used for LIF sampling \cite{kungl2018accelerated}, a spike-based implementation of Bayesian computing.
The hybrid plasticity scheme of BrainScaleS-2 has been succesfully applied in a maze runner task, where the neuromorphic agent has been trained in the \emph{learning-to-learn} framework \cite{bellec2018long,bohnstingl2019neuromorphic}.
Also using the plasticity processor, we have optimized spiking networks to task complexity by tuning the distance to a critical point \cite{cramer2019control}.
As a first implementation of reinforcement learning on BrainScaleS-2, a virtual player was trained in the game of \emph{Pong} \cite{wunderlich2019demonstrating}.

Here, we also want to consider a reinforcement learning task \cite{sutton2018}, making use of a wide range of the system's functionality, demonstrating the successful application of the design methods presented in this paper.
In reinforcement learning, an \emph{agent} interacts with its \emph{environment} and tries to maximize its expected future reward, obtained by the environment.
Especially, we consider a \gls{rstdp} learning rule in a pattern detection experiment.
\Gls{rstdp} is a three factor learning rule, combining reward information provided by the environment with \acrshort{stdp}-type correlation data.
The latter are used as eligibility traces to solve the credit assignment problem \cite{Fremaux2013}.

The agent $i$ accumulates the instantaneous reward $R_i$ given by the environment to obtain an \emph{expected reward}
\begin{equation}
	\langle R_i \rangle \leftarrow \langle R_i \rangle + \gamma (R_i - \langle R_i \rangle) \, ,
	\label{eq:expected_reward}
\end{equation}
where $\gamma$ scales the impact of previous trials.
Reward, mean reward, as well as the causal \gls{stdp} traces $e_{ij}$ enter the weight update equation
\begin{equation}
	\Delta w_{ij} = \eta \cdot (R_i - \langle R_i \rangle) \cdot e_{ij} + \xi_{ij} \, ,
	\label{eq:weight_update}
\end{equation}
with a learning rate $\eta$ and a random walk $\xi_{ij}$, and $j$ denoting the pre-synaptic neuron.

In the following we consider a task, where we stimulate a population of neurons via 16 input channels.
Each input emits Poisson distributed background spikes with a rate $\nu$.
Two patterns, termed A and B, are embedded into this noise floor (\prettyref{fig:overview_rstd}).
Each pattern consists of temporarily correlated spikes on five fixed input channels.
The two patterns can be configured to incorporate overlapping channels to increase task complexity.
In the course of the experiment, the network is trained such that all even neurons emit a spike when pattern A is applied, whereas all odd neurons fire when stimulated with pattern B.
In the case where no pattern is shown, all output neurons should remain silent.

An instantaneous binary reward $R_i$ is assigned to each neuron $i$:
In case a neuron fires succesfully according to the applied pattern -- or remains silent in absence of its specific pattern -- it obtains a reward $R_i = 1$.
If it, however, emits a spike when it is exposed to the opposite stimulus -- or only background noise --, it receives no reward, i.e. $R_i = 0$.

The update rule is implemented on the \gls{ppu}.
It reads out the neuronal rate counters in short intervals to determine the instantaneous success and assign reward signals.
Based on the latter, the expected reward is continuously updated in memory as a running average of the previously collected reward.
The processor furthermore reads synaptic correlation measured by the analog sensors in the synaptic circuits.
Joining reward and these eligibility traces, the weight update is calculated in a parallel fashion using the vector unit.
In addition, the \gls{ppu} simulates the ``environment''. 
This includes the generation of input patterns as well as background spikes.

In the model, synaptic weights $w_{ij}$ are not restricted to have either a positive or negative sign.
As the synapse drivers on the neuromorphic platform are implemented according to Dale's law \cite{dale1934} and hence are exclusively configurable to be excitatory or inhibitory, we join two synaptic rows with opposite sign to represent a single input.
The \gls{ppu} can transition between positive and negative weights by exclusively writing the absolute value of the weight to only the synapse carrying the appropriate sign.

\begin{figure}
	\begin{tikzpicture}
		\node[anchor=north west] at (0,0) {\input{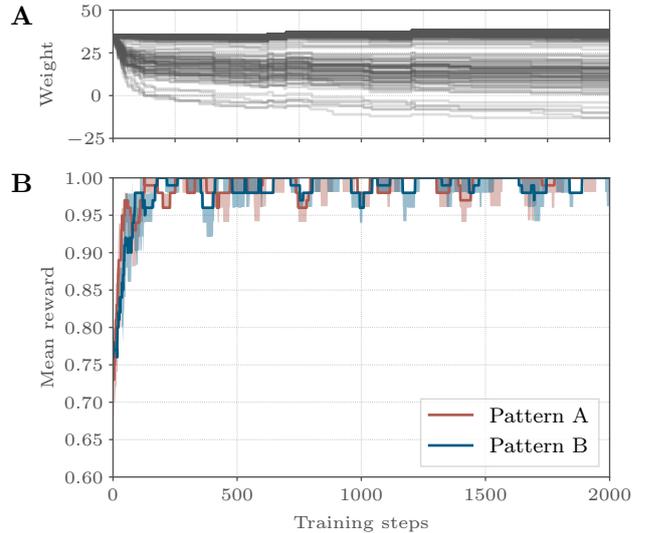}};
		\node at (0.2,-0.58) {\bfseries A};
		\node at (0.2,-2.80) {\bfseries B};
	\end{tikzpicture}
	\caption{%
		The mean expected reward converges to approximately one for all neurons during training.
		\textbf{A}~Weight evolution of all 256 synapses.
		\textbf{B}~Median mean reward reached by neurons in each population.
		The different colors correspond to the median mean expected reward inherent in neurons trained on pattern A (red) and pattern B (blue).
		Errors correspond to the 15 and \SI{85}{\percent} percentiles of the mean expected reward of the neurons in the respective population.
		All neurons reach a sufficiently high reward despite of pattern overlap.
	}
	\label{fig:mean_reward}
\end{figure}

The experiment was executed for 16 neurons on a BrainScaleS-2 prototype \cite{friedmann2016hybridlearning}.
For the results shown in \prettyref{fig:mean_reward}, the input patterns were overlapping by \SI{40}{\%}.
During training, the mean expected reward $\sum_i \langle R_i \rangle$ converges to approximately one for all neurons, indicating a state, where the neurons can discriminate between the two patterns.
The runtime of the experiment is heavily dominated by the transfer of firing rates and weight data to the host computer.
Reading out the experiment's state only at the end of training reduces the runtime per training step to \SI{290}{\micro\second}.
This demonstrates the advantages of a hybrid system combining an accelerated neuromorphic core with a flexible plasticity processor.

\section{Discussion}
\label{sec:discussion}
We presented implementation and verification methods that we have developed and applied while designing the \SI{65}{\nano\meter} BrainScaleS-2 ASICs.
Digital logic is rigorously verified using the framework presented in \prettyref{sec:veri-rtl}.
Besides unit testing, we apply a DPI-based testbench for full-chip integration testing.
It is directly interfaced to the BrainScaleS software stack, which allows for an efficient co-design and -verification of hardware and software.
This way, our chips can be utilized directly after commissioning of the hardware systems.

In section \prettyref{sec:veri-fc} we presented a framework for Python-based control and evaluation of analog circuit simulations.
\emph{Teststand} allows for the efficient implementation of pre-tapeout calibration algorithms, especially of interest in conjunction with \gls{mc} simulations.
This verification strategy has shown to dramatically increase in-silicon usability.
Leveraging the rich ecosystem of Python, the method is applicable to complex optimization tasks.
Circuits can easily be benchmarked against arbitrary datasets or even numerical simulations of a reference design.
It furthermore allows for the optimization of circuit designs themselves, e.g. by applying evolutionary algorithms to optimize transistor sizing.

Physical implementation of our ASICs is carried out using methodologies described in \prettyref{sec:phys-impl}.
A novel strategy for timing constraint derivation at design partition boundaries has been presented and applied to signals between \gls{ppu} partition and anncore.
Successful timing closure on this interface has been proven in silicon, albeit the overall target clock frequency of $\SI{500}{\mega\hertz}$ in the \gls{ppu} partition could not be reached due to a critical path that should be eliminated in a future chip revision.
Furthermore, a constraint strategy for the skew-minimized implementation of source synchronous signals to the event interface has been presented and verified in all process corners.
First measurement results, presented in \prettyref{sec:veri-fc}, of the STP circuits utilizing these event interfaces also prove a successful implementation.

Although the described methods for timing characterization and abstract generation, as well as the presented physical design methods should be applicable to similar problems, also outside the neuromorphic domain, the overall methodology is currently targeted at chips containing one anncore and up to two \glspl{ppu}.
When scaling the BrainScaleS-2 system up, it is conceivable to place several blocks combining anncore and two \glspl{ppu} on one full-sized reticle.
First, the methodology would have to be extended with an additional partitioning step for this block, accounting for the interface timing at the entry points of the routing channels in the anncore abstract.
Second, we are currently not applying dedicated techniques to reduce dynamic power in the digital logic, besides automated clock gating and the manually added clock gates.
To improve on this, more fine-grained autometic clock gating, and the frequency scaling features provided by the \gls{pll} \cite{hoeppner2013pll}, should be used for the design of larger systems.

To summarize the successful application of the methods described in this paper, we presented an experiment involving major parts of the BrainScaleS-2 hybrid plasticity architecture in \prettyref{sec:applications}.
With this example we hope to illustrate that a successful ASIC implementation of an accelerated analog neuromorphic system including a flexible and programmable plasticity scheme does not only rely on the circuit architecture but is facilitated by powerful implementation methodologies as well as simulation and verification strategies.

\section*{Acknowledgements}

The authors would like to give a special thank to A. Hartel for the timing characterization of the anncore. We thank S. Höppner and S. Scholze from the group \textit{Hochparallele VLSI-Systeme und Neuromikroelektronik} of C. Mayr from TU-Dresden for the PLL macro cell \cite{hoeppner2013pll} and SerDes macros \cite{scholze2012link} used in the two latest BSS-2 chip revisions, J. Weis and A. Leibfried for providing measurement data.

We especially express our gratefulness to the late Karlheinz Meier who initiated and led the project for most if its time.

\bigbreak

\noindent
This research was supported by the EU 7th Framework Program under grant agreements 269921(BrainScaleS), 243914 (Brain-i-Nets), 604102 (Human Brain Project) and the Horizon 2020 Framework Program under grant agreements 720270 and 785907 (Human Brain Project, HBP).

\bibliographystyle{abbrv}      %
\bibliography{vision}   %

\end{document}